\documentclass[useAMS,usenatbib]{mn2e}
\usepackage[T1]{fontenc}
\usepackage{ae,aecompl}
\usepackage{amsmath}
\usepackage{graphicx}
\usepackage{latexsym}
\usepackage{rotating}
\usepackage{float}
\title{Polarization morphology of SiO masers in the circumstellar
  envelope of the AGB star R  Cassiopeiae}

\author[K. A. Assaf et al.]{K. A. Assaf$^{1,2}$, P. J. Diamond$^{2,3,4}$,
A. M. S. Richards$^2$ and M. D. Gray$^2$\\
$^1$University of Wasit, College of Science, Dept. of Physics,  Kut, Wasit, Iraq\\
$^2$JBCA, Alan Turing Building, School of Physics and Astronomy, The
University of Manchester,  M13 9PL, UK\\
$^3$SKA Organisation, Jodrell Bank Observatory, Cheshire, SK11 9DL, UK\\
$^4$CSIRO Astronomy and Space Sciences, PO Box 76, Epping, NSW 1710, Australia
}

\begin{document}

\date{Accepted MNRAS 2013 February}

\pagerange{\pageref{firstpage}--\pageref{lastpage}} \pubyear{2010}

\maketitle

\label{firstpage}
\begin{center}
\begin{abstract}
   Silicon monoxide maser emission has been detected in the
   circumstellar envelopes of many evolved stars in various
   vibrationally-excited rotational transitions. It is considered a
   good tracer of the wind dynamics close to the photosphere of the
   star.  We have investigated the polarization morphology in the
   circumstellar envelope of an AGB star, R Cas. We mapped the linear
   and circular polarization of SiO masers in the $v=1$, $J=1-0$
   transition. The linear polarization is typically a few  tens of percent
   while the circular polarization is a few percent. The fractional
   polarization tends to be higher for emission of lower total
   intensity.  We found that, in some isolated features the fractional
   linear polarization appears to exceed 100$\%$. We found the Faraday
   rotation is not negligible but is $\sim$ 15$^\circ$, which could
   produce small scale structure in polarized emission whilst total
   intensity is smoother and partly resolved out. The polarization
   angles vary considerably from feature to feature but there is a
   tendency to favour the directions parallel or perpendicular to the
   radial direction with respect to the star.  In some features, the
   polarization angle abruptly flips 90$^\circ$. We found that our
   data are in the regime $g\Omega \gg R \gg \Gamma$, which indicates
   that the model of \citet{goldreich1973} can be applied and the
   polarization angle flip is caused when the magnetic field is at
   close to $55^{\circ}$ to the line of sight. The polarization angle
   configuration is consistent with a radial magnetic field although
   other configurations are not excluded.
\end{abstract}
\end{center}


\section{Introduction}
\label{s1}
The circumstellar envelope (CSE) in an asymptotic giant branch star
(AGB) is a very active region.  Any AGB star will lose mass in
the form of a slow wind at a rate that will significantly affect the
mass of the star as well as enriching the  interstellar
medium with nuclear processed material.
This produces a circumstellar envelope of escaping
dust and gas particles. In the zone just above the photosphere, out to
5$R_\star$ (stellar radius), the extended atmosphere experiences
periodic shocks, so each period is divided into intervals of mass
outflow and infall. In this zone, the cycle-averaged temperature drops
as the wind flows away from the star, from about 3100 K at the
photosphere to $\approx$ 750 K at 5.5$R_{\star}$ \citep{gray2009};
cooling takes place by line radiation from various molecules,
especially H$_2$O.  SiO masers provide sub-milliarcsec images of this region.

Some molecular species are formed in an
equilibrium process deep in the atmosphere and are destroyed in the
outer parts of the outflow by interstellar ultraviolet radiation
(H$_2$, CO, H$_2$O).  Other species, for example SiO, are depleted due
to condensation on dust particles at a few stellar radii
\citep{habing1996}.

 The SiO maser zone is thought to be shaped
by a combination of shocks from stellar pulsations, gravity and
possibly magnetic effects,
with radiation pressure on dust becoming possible towards the outer
region. SiO masers are seen in rings, with faint or no emission along
the line of sight to the star (\citealt{diamond2003};
\citealt{cotton2006}). This is due to
tangential beaming from an accelerating, approximately spherical wind,
where masers have the deepest amplification paths if they are at similar
velocities and in a similar plane of the sky to the star.

Studying the magnetic field properties
is done by investigating the polarization morphology of the maser
emission.
 The  maser emission is significantly linearly
polarized but circular polarization is weaker, as expected for a
non-paramagnetic molecule \citep{herpin2006}.
Fractional circular
polarization of SiO masers in late-type
stars from single-dish  observations is $m_c \le 0.5\%$
\citep{habing1996} while the degree of linear
  polarization is $m_\ell \sim 15 - 30\% $
  \citep{troland1979}.

There are two  explanations suggested for the linear polarization of
SiO masers.
Polarization can be produced by the magnetic field
(\citealt{goldreich1973}; \citealt{elitzur1992}).
 Alternatively, in the absence of a magnetic  field,  anisotropic pumping can
  produce  strongly polarized maser emission \citep{western1983}.
  This mechanism has been proposed as the cause of  the tangentially
  polarized maser emission
  seen in the map of SiO masers observed by the
  VLBA \citep{desmurs2000}.

If it is magnetic in origin, the linear polarization position angle
provides information about the structure of the magnetic field in
circumstellar envelopes.  The theoretical relationship between the
polarization position angle and the projected magnetic field direction
depends on $\theta_{\mathrm F}$, the angle between the magnetic field
direction and the maser propagation direction (our line of
sight).
When $\theta_{\mathrm F} <$ 55$^\circ$, the linear
polarization vectors are parallel to the magnetic field lines and when
$\theta_{\mathrm F} >$55$^\circ$ the vectors are perpendicular to the
field lines. When $\theta_{\mathrm F} \simeq$55$^\circ$ the linear
polarization vectors can flip $\frac{\pi}{2}$ within a single
feature. At this value of $\theta_{\mathrm F}$, the Van Vleck angle of
55$^\circ$, the masing region reaches maximum intensity and the
fractional linear polarization approaches zero (\citealt{elitzur1992};
\citealt{goldreich1973}).

R Cassiopeia is an oxygen-rich AGB star which is classified as an
M-type Mira-variable. Its optical brightness varies from magnitude
$+4.7$ to $+13.5$ with a period of 430 days and its mass is about 1.2
M$_{\odot}$. \citet{vlemmings2003} used astrometric  VLBI  to measure
a distance of $176^{+92}_{-45}$ pc with
a proper motion of (85.5 $\pm$0.8, 17.5 $\pm$0.7) mas yr$^{-1}$ in
R.\ A. and Dec., respectively. Various estimates of the stellar
velocity $v_{\star}$ appear in the literature, we have adopted the
value of 24$\pm$2 km s$^{-1}$.

We monitored the 43 GHz SiO masers of R Cas for almost two stellar
cycles. The total intensity results were discussed in
\citet{assaf2011}. Here, we describe the polarization observation in
section \ref{s2}. We present the polarization results in section
\ref{s3}, and
analyse these results in section \ref{s4}, drawing conclusions in
section \ref{s5}.

\section{Observations and Data reduction}{\label{s2}}

The SiO masers around R Cas were observed as part of a more extensive
programme of VLBA\footnote{The VLBA (Very Long Baseline Array) is
operated by the the National Radio Astronomy Observatory, a facility
of the National Science Foundation operated under a cooperative
agreement by Associated Universities, Inc.} monitoring of other
stars. In total we have 23 epochs of data for R Cas. Data were
recorded at each VLBA antenna in dual-circular polarization in two 4
MHz windows, each digitally sampled at the full Nyquist rate of 8 Mbps
in 1-bit quantization. The lower spectral window was centred at a
fixed frequency corresponding to the $v=1$, $J=1-0$ SiO transition, at an
assumed rest frequency of 43.12207 GHz and a systemic velocity
V$_{\mathrm {LSR}}$ = +24 km s$^{-1}$. R Cas was observed for three
45-minutes periods evenly spread over the 8 hour duration of the
run. Adjacent to each R Cas observation, 5 minutes was spent observing
the continuum calibrator 0359+509 at the same frequency as R Cas. The
data were correlated at the VLBA correlator in Socorro, NM. The
correlator accumulation interval was set to 2.88 seconds. All
polarization correlation products (RR, RL, LR, LL) were formed. This
configuration produced auto- and cross-power spectra in each 4 MHz
baseband with a nominal velocity spacing of $\sim$ 0.2
km.s$^{-1}$.

We reduced the data using the standard approach to VLBI spectroscopy
within the NRAO {\sc aips} package
(http://www.aips.nrao.edu/cook.html). We processed the visibility data
using the the semi-automated spectral-line polarization calibration
pipeline which was originally written by Diamond in 1998, based on the
formalism in \citet{kemball1995} and developed further by
\citet{kemball1997} (for more detail see \citealt{assaf2011}).

 After the pipeline processing, we noticed that
there were systematic offsets in flux density of a few to 15\% between
LL and RR spectra for each epoch.  We assumed that the net circular
polarization of SiO across the band is zero and the total intensity
calibration is correct. This enabled us to calculate corrections to
align the flux scales for LL and RR and obtain realistic values of
Stokes $V$, albeit with large uncertainties.

 The absolute Electric Vector Position Angle ({\sc evpa}) for any linearly
 polarized emission is unknown because there is no instrumental
 measurement of absolute R-L phase difference in the reference antenna
 of the VLBA and in the VLA, but the VLA can provide absolute
 astronomical calibration of {\sc evpa} relative to a small subset of
 primary astronomical calibrators, given the lower spatial resolution
 of the array. We used the VLA to transfer absolute polarization
 calibration to VLBA data via the compact secondary polarization
 calibrator (0359+509) which was observed by both arrays on 19$^{\mathrm{th}}$
 Dec 1999. The same reference antenna {\sc kp} was used in the
 reduction of all epochs of VLBA data and the polarization angle
 stability is likely to be better than 35$^\circ$ (C. Walker private
 communication).
 Hence, we were  able to estimate the polarization angle of 0359+509 as $30^{\circ} \pm 10^{\circ}$.

We fitted 2-D Gaussian components to each patch of total intensity
(Stokes $I$) maser emission brighter than $5\sigma_{\mathrm {rms}}$, to
measure the position and flux density, as described in \citet{assaf2011}.  We
measured the $Q$, $U$ and $V$ flux densities at the position of the
peak of the $I$ component, as described in \citet{kemball2009}. We
use the peak (rather than integrated) intensity measurements of Stokes
$I$ for comparison with the other Stokes parameter measurements.  We
estimated the uncertainties based on (beam size)/(signal-to-noise
ratio), as appropriate for a sparse array (\citealt{condon1998},
\citealt{richards1999}).

We formed the linearly polarized intensity
\begin{equation}
P=(Q{^2}+U{^2})^{1/2}
\label{E1}
\end{equation}
 the fractional circular polarization
\begin{equation}
 m_c=\frac{V}{I}
\label{E2}
\end{equation}
 the fractional linear polarization
\begin{equation}
m_{\ell}=\frac{P}{I}
\label{E3}
\end{equation}
and the electric vector polarization angle {\sc evpa}

\begin{equation}
 {\small{\mathrm{EVPA}}}=0.5\mathrm{arctan}\frac{U}{Q}
\label{E4}
\end{equation}
(e.g. as defined by \citealt{heiles2002} or \citealt{elitzur1992}).

We measured the angular FWHM ($s$) of the total intensity components by
deconvolving the restoring beam from the measured size.  This was used
to estimate the brightness temperature $T_{\mathrm B}$ from the
integrated intensity.
Series of components form features and we estimated the average positions,
angular size $L$, angular FWHM $d_{\mathrm F}$, peak flux densities and other
properties of each feature.

The largest angular separation between components making up a feature,
$L$, measures the actual angular size of the emission detected. The
separation of the components with flux density closest to half the
peak, $d_{\mathrm F}$, represents the beamed size of the feature, as
explained in \citet{richards2011}, with reference to Elitzur (1992).
The beaming angle $d\Omega=(\frac{d_{\mathrm F}}{L})^2 $.  We note
that $d_{\mathrm F}$ and $L$ may be underestimated, as figs B1-3 in
\citet{assaf2011} show that a significant amount of total intensity
flux is resolved out. There is no clear, systematic difference between
the effect on brighter or fainter emission, so we suggest that the
main effect is to increase the uncertainty in $d\Omega$, but it could
be somewhat overestimated if the fainter emission is more diffuse,
leading to $L$ being more underestimated.

An improved method of
finding the centre of emission has led to some minor changes with
respect to \citet{assaf2011}, which do not affect those results
significantly. Table (\ref{t1}) shows the updated values of
the average shell radius and the shell thickness, and the difference
with the old values.
The differences are close to the uncertainties
(given in  \citealt{assaf2011}), which are higher for the last few
epochs due to the fainter emission and large gaps in the maser shell.
\begin{table}
\begin{center}
\begin{tabular}{|c|c|c|c|c|} \hline
 Epoch  & New ($R$) &error & d$\emph{r}$       &Abs(diff)\\ \hline
 BD62A  &25.93 &  0.67    &8.72          &1.16\\
 BD62B  &25.49 &  0.05    &9.05          &1.59\\
 BD62C  &24.72 &  1.77    &8.21          &1.33\\
 BD62D  &24.24 &  1.22    &8.60          &1.81\\
 BD62E  &24.86 &  1.59    &7.87          &2.24\\
 BD62F  &25.45 &  1.98    &5.80          &2.98\\
 BD62G  &25.96 &  2.11    &6.20          &1.45\\
 BD62H  &30.34 &  0.30     &7.92          & 2.25\\
 BD62I  &27.50 &  0.53    &5.38          &0.99\\
 BD69A  &26.77 &  1.64    &7.52          &0.52\\
 BD69B  &27.97 &  0.37    &6.19          &0.26\\
 BD69C  &27.55 &  1.50     &5.68          &  0.45\\
 BD69D  &28.09 &  0.16    &5.51          & 0.11\\
 BD69E  &28.02 &  0.50     &6.07          &  0.52\\
 BD69F  &28.32 &  1.71    &6.11          & 0.17\\
 BD69G  &24.24 &  2.50     &8.23          &  4.28\\
 BD69H  &29.99 &  0.26    &9.62          &0.83\\
 BD69I  &24.90 &  0.17    &7.68          & 2.25\\
 BD69J  &24.60 &  0.63    &14.05         & 2.23\\
 BD69K  &21.41 &  3.11    &13.51         & 3.70\\
 BD69L  &18.17 &  1.22    &4.50          &5.79\\
 BD69M  &17.27 &  3.83    &7.82          & 8.60\\
 BD69N  &21.97 &  1.68    &11.21         & 4.85\\

     \hline
\end{tabular}
\caption{The new shell radius (New $R$),the error in $R$, the shell thickness
  (d$r$) and the absolute difference between the old and the
  new values, all in mas.}
\label{t1}
\end{center}
\end{table}

\section{Results}{\label{s3}}
We made  Stokes $I$, $Q$,
$U$ and $V$ image cubes of R Cas, at each epoch, at spatial and
spectral resolutions of
approximately ($40\times20$) $\mu$as$^2$, 0.2 km s$^{-1}$.

The polarization
detection threshold for individual components is
$5\sigma_{\mathrm{rms}}$ and $m_{\ell} >5$\% or $m_c >15$\%; lower
thresholds are possible when averaging over larger spectral or spatial
regions (including correction for Ricean bias). Each feature is made up of many components, and every feature contain more than 0.5 Jy/beam summed total intensity. Quiet channels have $\sigma_{\mathrm {rms}} 0.03 - 0.04$ Jy beam$^{-1}$, so any linear polarization with $m_{\ell} \ge 20$ percent will also have a signal to noise ratio $\ge5$.

\subsection{Circular Polarization}
The mean degree of circular polarization for each epoch was estimated
using Eq.~\ref{E2}.

We found the fractional circular polarization is $m_c \sim 0.4 \to 6
\%$. We attempted to fit the expression given by \citet{elitzur1996}
to the first derivative of Stokes $V$. This
failed due to the weakness of  Stokes $V$ intensity and the large channel
separation which is greater than the Zeeman splitting.

\subsection{Linear Polarization}
\label{sec:linpol}

\begin{figure*}
  \includegraphics[height=230mm]{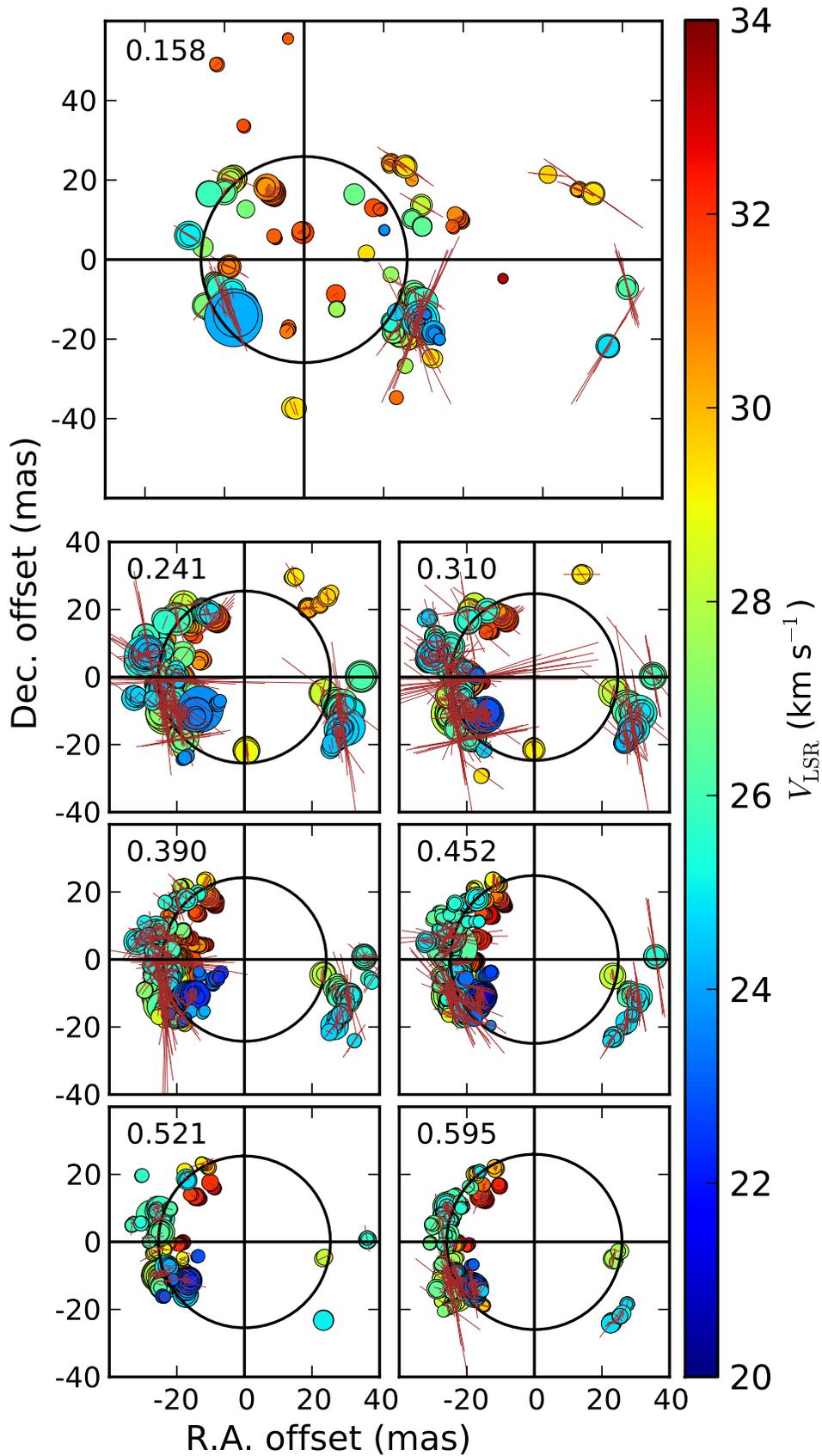}
\caption{Polarization morphology of R Cas SiO masers. Each pane is
 labelled with the stellar phase. Each symbol represents an
 individual, fitted maser component. Symbol area in mas$^{2}$
 represents 1/8 of total intensity in Jy. The vectors show the orientation
 of the {\sc evpa} and 5 mas in length represents $P$ 1 Jy
 beam$^{-1}$.  The black circles show the weighted shell radii given
 in Table~\ref{t1}. Epochs (1-7)}
\label{f1}
\end{figure*}

\begin{figure*}
\includegraphics[height=230mm ]{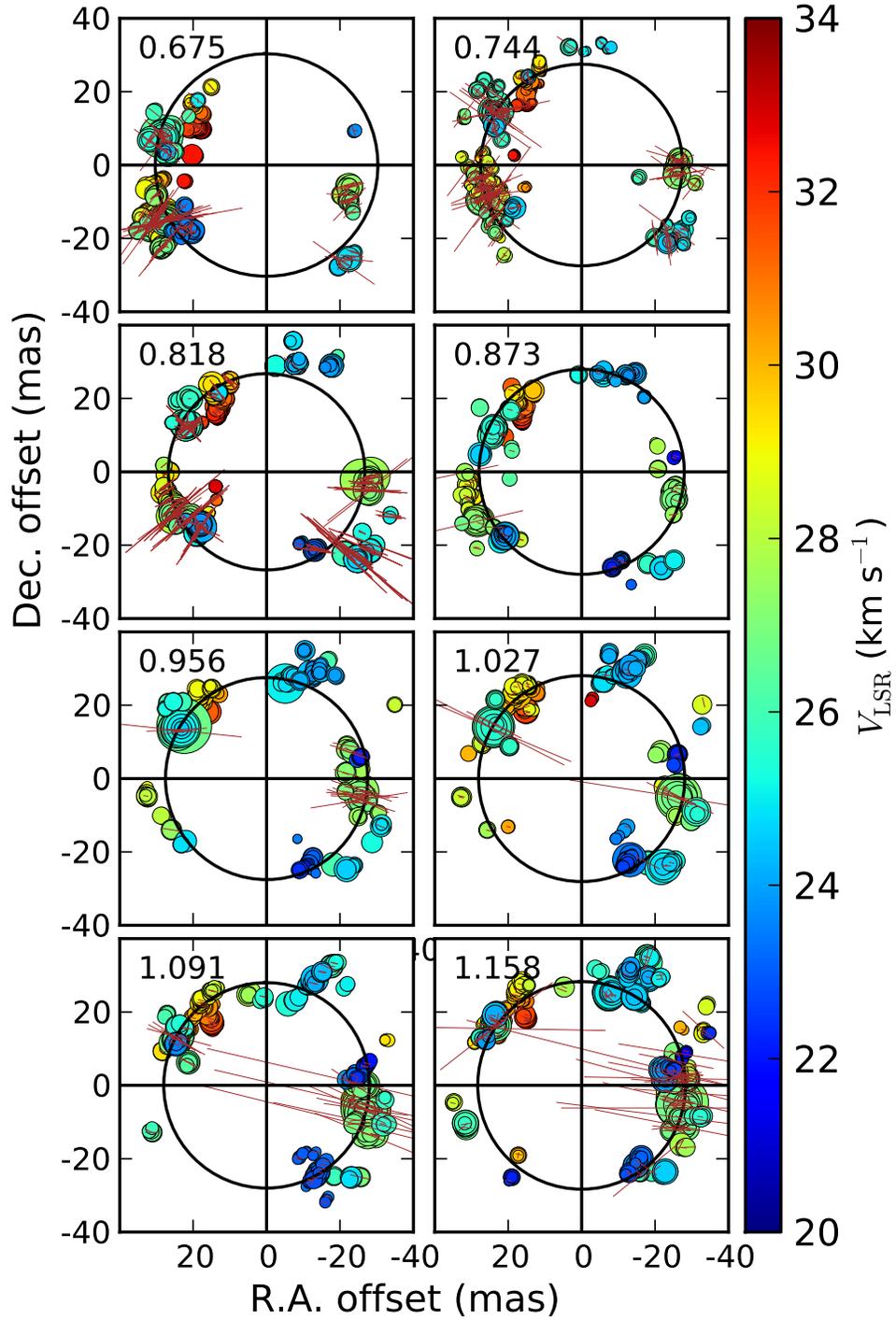}
\caption{Same as Fig\ref{f1} but for Epochs 8-15}
\label{f2}
\end{figure*}

\begin{figure*}
\includegraphics[height=230mm ]{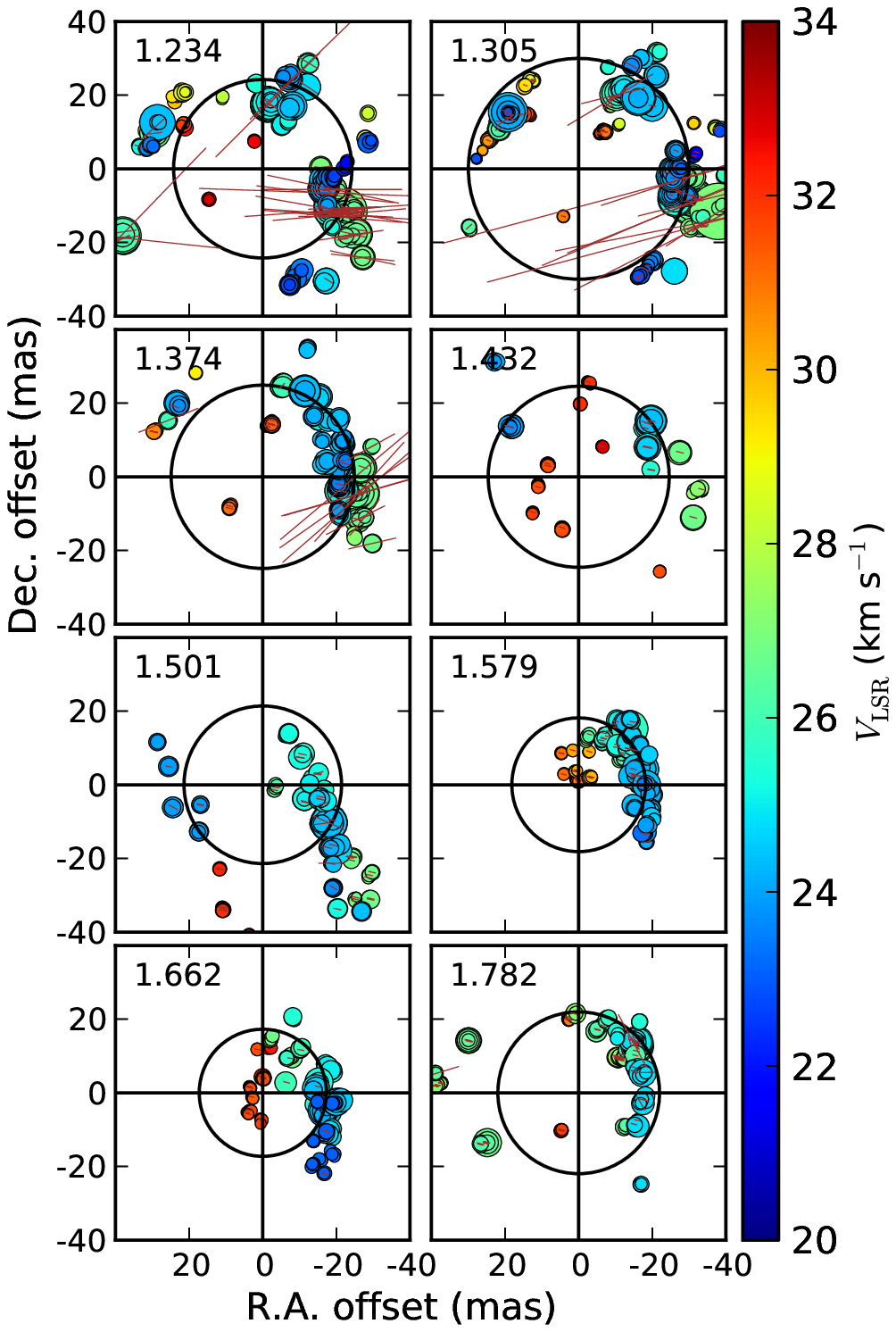}
\caption{Same as Fig\ref{f1} but for Epochs 15-23}
\label{f3}
\end{figure*}

\begin{figure}
  \includegraphics[width=0.5 \textwidth ]{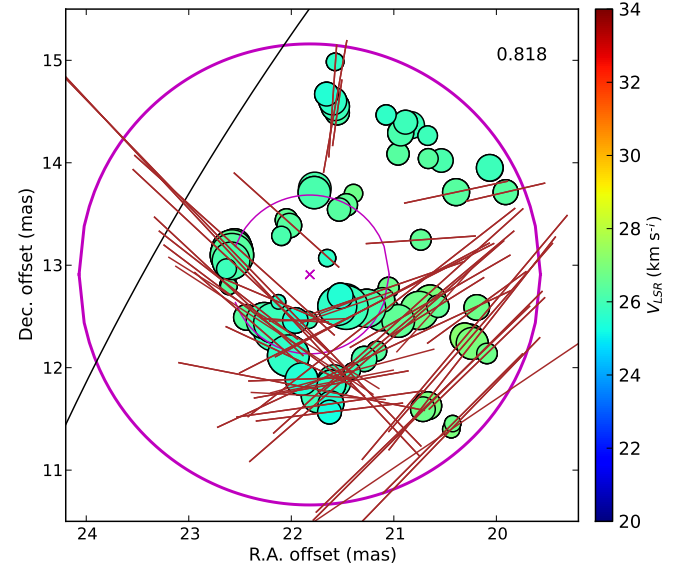}
\caption{Enlargement of one feature from the NE of the maser shell imaged at $\phi=0.818$, from Fig.~\ref{f2}.
The vectors show the orientation of the {\sc evpa} and 2
 mas in length represents  $P$ 1 Jy beam$^{-1}$.  The inner and outer
 magenta circles have diameters corresponding to $d_{\mathrm F}$ and
 $L$, respectively. 
}
\label{RCAS2_newpol818.eps}
\end{figure}

Figs. \ref{f1}, \ref{f2} and \ref{f3} show the polarization morphology of
SiO masers in R Cas
for 23 epochs. The linear polarization vectors are superimposed on
the map of the total intensity. The orientations of the vectors
define the electric field plane and the length of the vectors
are proportional to the linearly polarized intensity
$P$.  Fig.~\ref{RCAS2_newpol818.eps} shows an enlargement of the
components making up a single feature.  The magenta cross marks the
error-weighted centroid of the feature. The thin magenta
circle encloses half the feature flux, i.e. its diameter is the
feature FWHM, $d_{\mathrm F}$ and the diameter of the thicker magenta
circle represents the total feature extent $L$. 

 We found that SiO maser emission is
significantly linearly polarized. We calculated the fractional linear
polarization using Eq. \ref{E3}
We found that the percentage of
linear polarization (averaged over all features per epoch) is $m_{\ell}\sim
11 \to 58 \%$  but exceeds 100\% in
some isolated features, discussed in Section~\ref{sec:fpolhi}.

\begin{figure}
\includegraphics[width=0.5 \textwidth ]{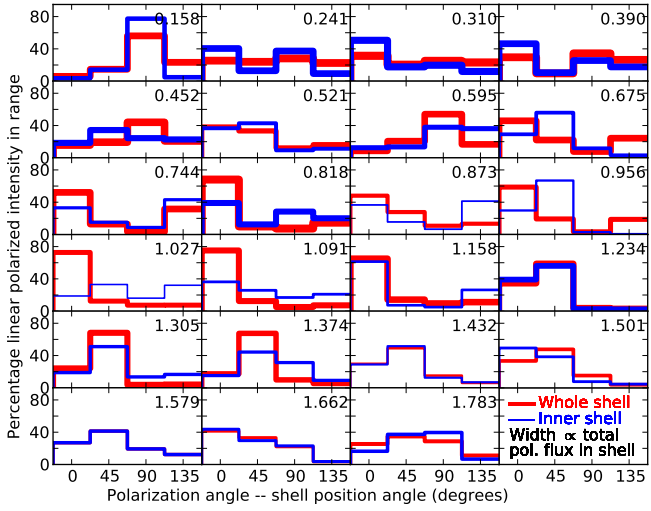}
\caption{ Histogram of the distribution of polarized emission with
  respect to the deviation of the polarization angle from the radial
  direction ({\sc evpa}$-\theta$). The line thickness is proportional
  to the logarithm of total polarized emission intensity (at all
  angles). The maxima occur at $\phi=0.390$, reaching 160 Jy in the
  inner shell and 350 Jy in the whole shell. The minima occur at
  $\phi=1.432$, falling to 4 Jy in the inner shell and 7 Jy in the
  whole shell.}
\label{f4}
\end{figure}

We investigated the relationship between the {\sc evpa} and the radial
direction with respect to the star, defined by the position angle in
the plane of the sky, $\theta$. Fig.\,\ref{f4} shows the proportion of
the polarized emission within bins of ({\sc evpa}$-\theta$) in the
whole SiO shell and in the inner shell (within the radius enclosing
25\% of the total maser flux at each epoch). The thickness of the line
is proportional to the logarithm of the total linearly polarized
flux. During the first cycle ($\phi=0.158$ to $\phi=1.158$, 15
epochs), we found that the polarized flux in the whole shell is
dominated by emission with {\sc evpa}  parallel to  $\theta$ at 10 epochs. The
most popular {\sc evpa} orientation with respect to the position angle
$\theta$ is perpendicular for the other 5 epochs, dominating in 3 of
these. However, in the inner shell, only half the epochs have a high
proportion of emission with parallel (6 epochs) or perpendicular (2
epochs) {\sc evpa}; it is at intermediate angles for the remaining 7
epochs of the first cycle.  In the second cycle, 7 out of the 8 epochs
(from $\phi=1.234$ to $\phi=1.783$), are dominated by emission with
intermediate {\sc evpa} in the whole shell but this comes mainly from
a single feature. At $\phi= 1.662$ the {\sc evpa} is predominantly
parallel. The inner shell has more emission with a parallel {\sc evpa}
at epochs $\phi=1.501, 1.662$, otherwise behaving similarly to the
whole shell.

The polarization structure during the first stellar cycle can be
summarized as a bimodal distribution. Most of the linear polarization
vectors are either radial (parallel to the position angle of the
location of the emission in the projected shell) or tangential
(perpendicular). However, the polarization in the inner part of the
shell is somewhat less ordered.  The later parts of the second stellar
cycle do not show any clear pattern, but the emission generally was
noisier with fewer significantly polarized features.

\section{Discussion}{\label{s4}}

We noted in Section~\ref{s1} that there are two competing models put
forward to explain SiO polarization.
 The adopted  model for the  transfer of the polarized
 maser emission in a non-paramagnetic molecular transition  plays an
 important role in the interpretation of the polarization measurements
 \citep{elitzur1996}.

First, we made an independent estimate of the magnetic field strength
to test the feasibility of the magnetic model.  The bulk kinetic
energy density ($E_{\mathrm{Bulk}}=\frac{1}{2}\rho v^2$) is
$\le2.1\times10^{-3}$ J m$^{-3}$, where $\rho$ is the volume density
of SiO, taken as 1.42$\times$10$^{-13}$gm cm$^{-3}$ \citep{gray2009},
and $v$ is the expansion velocity which is up to 5.5 km s$^{-1}$
\citep{assaf2011}. The thermal energy density
$E_{\mathrm{Thermal}}=\frac{3}{2}n_{\mathrm H_{2}}kT$ is $0.832\times10^{-3}$ J m$^{-3}$,
where $n_{\mathrm H_2}$ is the molecular hydrogen  density, $k$ is Boltzmann's
constant and $T$ is the effective temperature, 1330 K
\citep{gray2009}. A first estimate of $B$ can be found by comparing the
energy densities. The magnetic energy density $E_\mathrm B$ is given
by
\begin{equation}
E_\mathrm B=\frac{B^2}{2\mu_\circ}\sim
4\times10^{-3}\left(\frac{B}{{\mathrm G}}\right )\; {\mathrm {J\; m^{-3}}}
\label{E5}
\end{equation}
By equating these energy densities, we found the first estimate of B
$\sim$ 0.725 G. \citet{herpin2006} measured a magnetic field
strength  in the range  0.9 $\to$ 2.8 G from single dish observations of R Cas at 86 GHz, slightly higher than our estimate.

\subsection{Application of Zeeman interpretation}\label{Zeint}

We can decide which interpretation to apply by comparing three
parameters: the stimulated rate emission $R$, the Zeeman
splitting rate $g\Omega$ and the decay rate $\Gamma$.

We calculated the maser beaming angle $d\Omega $ and the maser
brightness temperature $T_\mathrm B$ from our data. We found the
error-weighted beaming angle (Table ~\ref{t2}) averaged over 23 epochs
is $d\Omega\sim$ 0.047 sr and the mean brightness temperature
$T_{\mathrm B}\sim$ 2.76 $\times 10^{9}$ K.
 The stimulated emission rate $R$ for saturated
masers in this transition is given by \citet{kemball2009}:
\begin{equation}
R=23 \left(\frac{T_{\mathrm B}}{2 \times 10^{10}
  {\mathrm K}}\right)\left(\frac{d\Omega}{10^{-2} {\mathrm sr}}\right) \; \mathrm{s}^{-1}
\label{E6}
\end{equation}
giving $R \sim$ 15 s$^{-1}$.

The Zeeman splitting rate for SiO is given by
$\frac{g\Omega}{2\pi}\simeq 200 B$ s$^{-1}$
\citep{plambeck2003}. Using the value of $B$ estimated above,
$g\Omega \sim$ 900 s$^{-1}$.
For this work, the Zeeman splitting ratio is greater than the
stimulated emission rate ($g\Omega\gg R$). This result excludes
intensity-dependent non-Zeeman circular polarization
\citep{nedoluha1994}. The radiative decay rate $\Gamma$ for the $v=1$,
$J=1-0$ transition is $<$ 1 s$^{-1}$ \citep{kemball2009}.
Our data are in the limit g$\Omega \gg R \gg \Gamma$, which indicates that
the \citet{goldreich1973} solutions for the linear polarization are
applicable. A field of this magnitude
($\sim 1$ G) also effectively rules out anisotropic radiative pumping
models on the grounds that the Zeeman precession rate is sufficient
to mix the magnetic sublevel populations efficiently. A similar
conclusion has also been reached by \citet{asensioramos2005}.

The ratio of $\frac{R}{\Gamma}$ indicates the saturation
level of the maser. Equation 8.6.2 of \citet{elitzur1992} shows that,
for our SiO transition, the masers are saturated when the brightness
temperature $T_{\mathrm B} > 0.5\times 10^9$ K, confirming that most
of the R Cas SiO masers are saturated.
We therefore interpret our results using the magnetic field model for
the origin of linear polarization.

Faraday rotation is proportional to the square of the wavelength
($\lambda^{2}$), and so it is smaller at
higher frequencies.  However, Faraday rotation is also proportional to the
electron density, which is higher at the region close to the star,
than at higher distances where the longer-wavelength masers originate.
Previously, it has been presumed that  Faraday rotation is
negligible for wavelength 7 mm as in our case. We used the relation
below from \citet{garcia1988}:

\begin{equation}
\psi_F=0.5\times\left(\frac{n_e}{{\mathrm{cm}}^{-3}}\right)
\left(\frac{B_{||}}{\mathrm{mG}}\right)
\left(\frac{L}{10^{15}{\mathrm{cm}}}\right) \left(\frac{\lambda}{18 {\mathrm{cm}}}\right)^2
\label{E7}
\end{equation}
where $n_e$ is the electron density and $L$ is the path length in the
maser region (maser shell thickness). From \citet{reid1997} we found
that the electron density in the SiO maser region $\sim$ 1500
cm$^{-3}$, hence the Faraday rotation is about 16$^\circ$ for $L\sim$
2$\times$ 10$^{13}$ cm for a magnetic field of $B\sim$ 750 mG. This
estimate is in fact slightly smaller than the estimates by
\citet{herpin2006}, so Faraday rotation could be somewhat higher. It
is still within the uncertainties of our {\sc evpa} measurements.

 \subsection{Linear polarization approaching or exceeding total intensity}
\label{sec:fpolhi}

\begin{figure}
  \includegraphics[width=0.5 \textwidth, height=0.45 \textwidth]{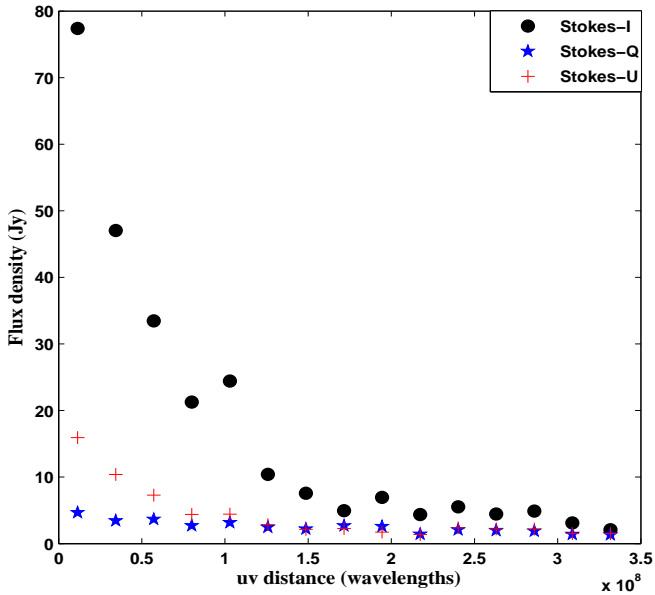}
\caption{ Stokes $I$, $Q$
 and $U$ flux density as a function of baseline length for channels
 averaged from 26.3 -- 27.4 km s$^{-1}$, $\phi=1.579$.}
\label{f5}
\end{figure}

We found that in some isolated features, the percentage of linear
polarization is greater than $100\%$ at three epochs $\phi = 0.744$,
$\phi = 1.158$ and $\phi = 1.305$,when it reaches 445\%. This is
likely due to resolving out of the
total intensity Stokes $I$ even on the shortest VLBA
baselines. Emission on scales greater than $\sim$ 5 mas cannot be imaged
by the VLBA. Comparison of the auto- and cross-correlation spectra
(\citet{assaf2011}, Figs B1-3) shows that 10--90\% of the total
intensity emission of R Cas is resolved out. Fig~\ref{f5} compares the
flux density as a function of baseline length for Stokes I, Q and U
for the spectral range including the feature with m$_\ell >$100$\%$ at
$\phi$=1.305.  Other features in the same channels have lower m$_\ell$
but the average visibility amplitudes still show that the total
intensity emission rises on large scales (shortest baselines) to a
much greater extent than polarized emission.
 The continuum calibration source 0359+509 does not show
any such effect.

More intense polarized than total intensity has
been observed before, in low-frequency continuum observations using WSRT
due to Faraday rotation creating smaller-scale
structure in the polarized emission \citep{haverkorn2003}.
 In R Cas, this could be due either to
fluctuations in the magnetic field, or to inhomogeneities in
the ionisation fraction, if these are on smaller scales than
turbulence in the neutral medium.

We examined the relationship between auto- and
cross-correlation spectra for other epochs of high fractional
polarization, for example $m_{\ell} \sim 70-80$\% around 27 km
s$^{-1}$ at $\phi = 1.158$ and 1.305
(Fig.~\ref{f10}). \citet{assaf2011}, figs B2-3 shows that about 50\%
of the total intensity flux is recovered in this velocity range at
$\phi = 1.027-1.432$. This suggests that the `true' fractional
polarization could be no greater than 30--40\% (including
uncertainties). A more exact comparison is not possible due to lack of
spatial resolution in auto-correlation data.

\begin{table}
\begin{center}
\begin{tabular}{|c|c|c|c|c|} \hline
Epoch  &   $\phi$     &  $d\Omega$ sr    & $T_{\mathrm{B}}$ K&m$_\ell$\\ \hline
      BD62A & 0.158  &0.066     & 1.75$\times10^{9}$&22.15  \\
      BD62B & 0.241  &0.044     & 0.96$\times10^{9}$&24.36 \\
      BD62C & 0.310  &0.039     & 0.90 $\times10^{9}$&35.65 \\
      BD62D & 0.390  &0.041     & 0.97 $\times10^{9}$&30.10  \\
      BD62E & 0.452  &0.039     & 1.18 $\times10^{9}$&30.66 \\
      BD62F & 0.521  &0.044     & 1.25 $\times10^{9}$&19.40  \\
      BD62G & 0.595  &0.028     & 0.38 $\times10^{9}$&44.97 \\
      BD62H & 0.675  &0.030     & 0.75 $\times10^{9}$&52.71 \\
      BD62I & 0.744  &0.048     & 1.78 $\times10^{9}$&57.91 \\
      BD69A & 0.818  &0.090     & 1.14 $\times10^{9}$&36.82 \\
      BD69B & 0.873  &0.030     & 0.13 $\times10^{9}$&11.13 \\
      BD69C & 0.956  &0.083     & 12.15 $\times10^{9}$&16.47 \\
      BD69D & 1.027  &0.041     & 1.27 $\times10^{9}$ &15.74 \\
      BD69E & 1.091  &0.022     & 5.13 $\times10^{9}$ &21.02\\
      BD69F & 1.158  &0.043     & 0.22$\times10^{9}$ &24.24\\
      BD69G & 1.234  &0.023     & 2.38 $\times10^{9}$&16.04\\
      BD69H & 1.305  &0.029     & 1.02 $\times10^{9}$&23.88 \\
      BD69I & 1.374  &0.052     & 3.89 $\times10^{9}$&21.14  \\
      BD69J & 1.432  &0.040     & 1.08 $\times10^{9}$&16.84  \\
      BD69K & 1.501  &0.051     & 3.57 $\times10^{9}$&16.09  \\
      BD69L & 1.579  &0.097     & 1.19 $\times10^{9}$&20.44 \\
      BD69M & 1.662  &0.062     & 0.45 $\times10^{9}$&23.19  \\
      BD69N & 1.783  &0.032     & 19.85 $\times10^{9}$&25.84   \\ \hline
\end{tabular}
\caption{Stellar phase $\phi$, error-weighted maser beaming angle
  $d\Omega$, error-weighted maser brightness temperature
  $T_\mathrm{B}$ and mean fractional linear polarization m$_\ell$.}
\label{t2}
\end{center}
\end{table}

\subsection{90$^\circ$ change in polarization angle}

\begin{figure}
\includegraphics[width=0.5 \textwidth]{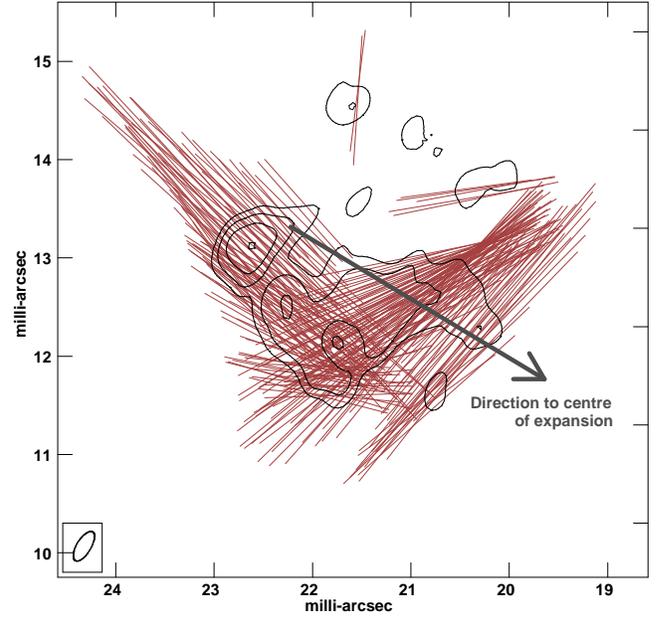}
\caption{The NE maser feature seen at $\phi=0.818$ in Fig.~\ref{f2} and enlarged in   Fig.~\ref{RCAS2_newpol818.eps}.  The mapped emission was summed
   over $V_{\mathrm{LSR}}$ 25.67 -- 27.43 km.s$^{-1}$. The contours are at (--1,1,2,4,8) $\times$ 1.7 Jy beam$^{-1}$. The vectors show the {\sc evpa}, plotted every two pixels (0.1 mas), length 1 mas = 0.8 Jy beam$^{-1}$ linearly polarized intensity. }
\label{f7}
\end{figure}

\begin{figure}
\includegraphics[width=0.5 \textwidth]{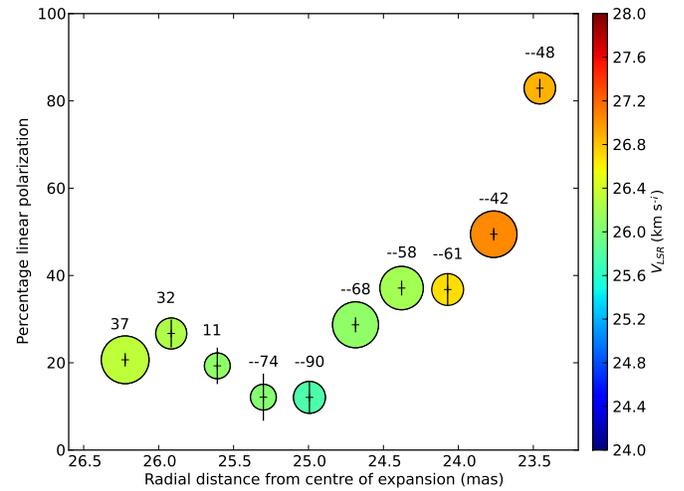}
\caption{The feature shown in Fig.~\ref{f7}.  The error-weighted mean percentage polarization is plotted for ten increments in radial distance from the centre of expansion ($r$).  Symbol size is proportional to the mean polarized intensity in each bin, from a minimum of 3 Jy around $r$ = 25.5 mas, to 11 Jy at $r$ = 23.8 and 26.2 mas.  The symbols are labelled with the error-weighted mean polarization angle. Note the expanded velocity scale.}
\label{RCAS2_polrad818.eps}
\end{figure}

 The linear polarization image of the north-east component in
 Fig~\ref{f7} shows an abrupt transition in {\sc evpa} (Electric Vector
 Position Angle) of approximately $\frac{\pi}{2}$ near $V_{\mathrm{LSR}}\sim
 26.78$ km s$^{-1}$. At this point, the linearly polarized intensity
 is near its minimum. The orientation of the polarization vectors is
 tangential to the projected ring in the part of the feature closer to
 the centre of expansion, whilst they are radial further out.
Fig.~\ref{RCAS2_polrad818.eps} shows the fractional linear polarization for this feature as a function of distance from the centre of expansion.  Note the expanded velocity scale. We ordered the components by increasing radial distance from the centre of expansion $r$ and found the error-weighted mean linearly polarized and total flux densities in ten bins in order to calculate the fractional polarization, plotted as a function of $r$.  Each point is labelled with the mean {\sc evpa} and symbol size is proportional to the polarized intensity.  The emission at $r<25$ mas has a mean {\sc evpa} $\sim-55^{\circ}$ and the emission at $r>25.5$ mas has a mean {\sc evpa} $\sim+30^{\circ}$.
The fractional polarization passes through a minimum ($\sim10$ percent) at $r\sim25-25.5$ mas, where the mean {\sc evpa} passes through the transitional angles of --74$^{\circ}$ to --90$^{\circ}$. The linearly polarized intensity is lower in this range, compared with most emission at more extreme {\sc evpa} values.

This is consistent with the prediction of
 \citet{goldreich1973} when the angle $\theta_{\mathrm F}$ between the
 magnetic field and the line of sight changes from $<55^{\circ}$ to
 $>55^{\circ}$, if the magnetic field is radial with respect to the
 star.  A similar phenomenon in TX Cam was described by
 \citet{kemball2011} and in W43A by \citet{vlemmings2006}.

\begin{figure}
  \includegraphics[width=0.51 \textwidth]{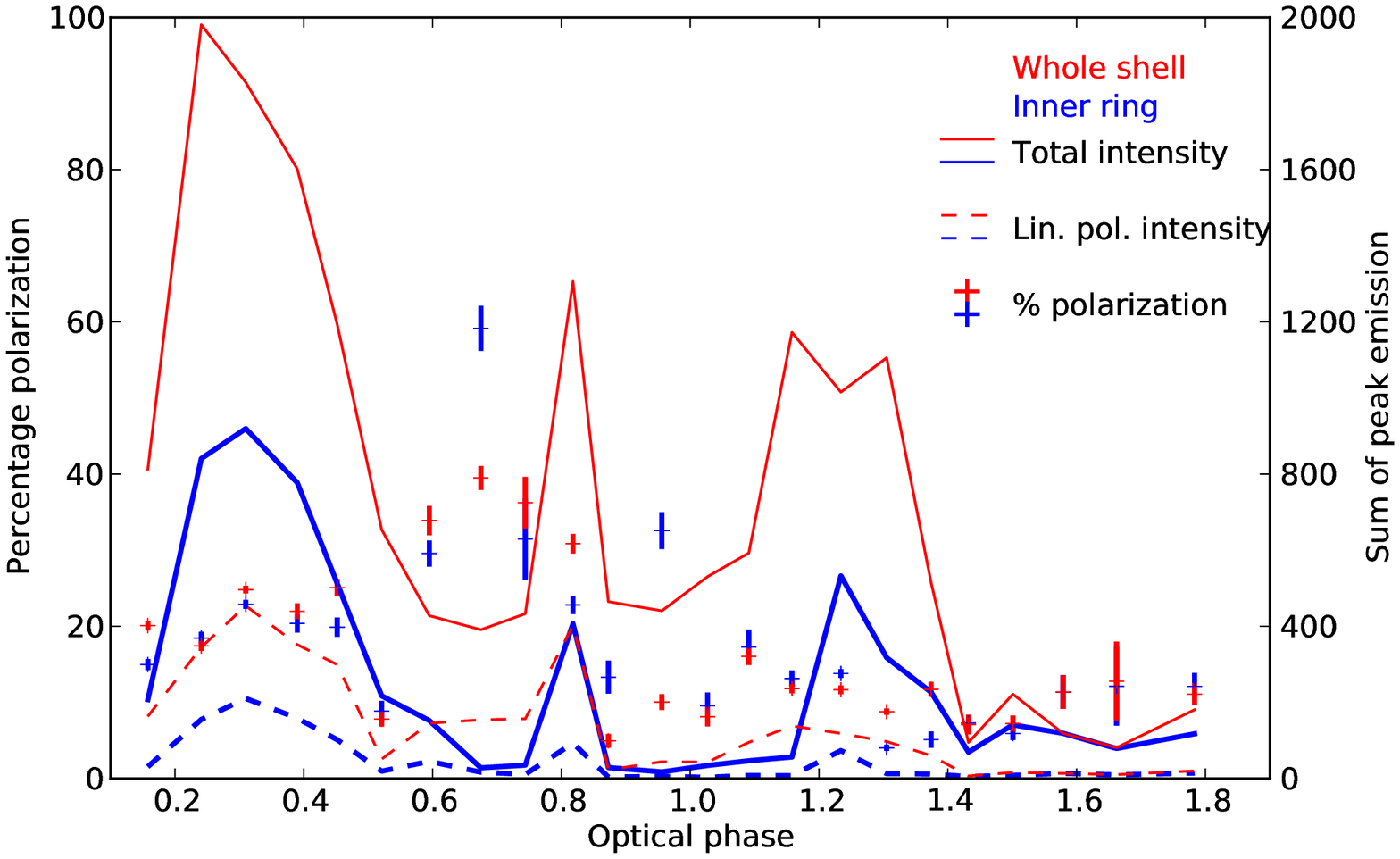}

\caption{Distribution of polarized intensity with respect to  epoch
in the inner ring and whole shell}
\label{f6}
\end{figure}

\subsection{Time variability of polarization}

Fig.~\ref{f6} shows the relationship between linear polarization and
total intensity in  the whole maser shell and in the inner ring
(radius 25 mas).  There is an anticorrelation between the total
intensity peak and the
fractional polarization during the first optical cycle. The percentage
flux density is similar or slightly higher in the whole shell compared
with the inner shell up to $\phi$= 0.595.
The highest fractional polarization, at $\phi$= 0.744, comes mainly
from the inner shell.  The fractional polarization is also higher in the
inner shell from $\phi$= 0.873 to $\phi$= 1.234.  At later epochs, the whole
shell has higher polarization but at epoch 1.432 and later, most of
the emission comes from the inner shell.

\begin{figure}
\includegraphics[width=0.55 \textwidth, height=0.45 \textwidth]{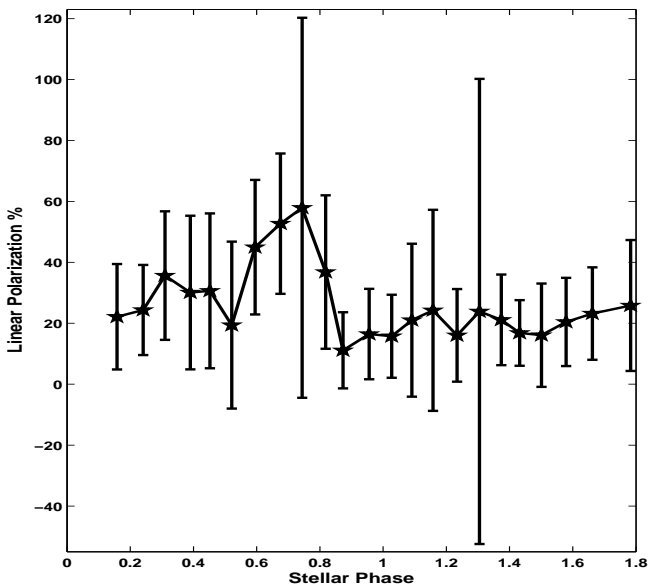}

\caption{The relationship between the fractional linear polarization and
  the stellar phase. The error bars represent the standard deviation.}
\label{f7.5}
\end{figure}

We investigated the variation of mean linear polarization  as a
function of stellar pulsation phase in
Fig.~\ref{f7.5}. We calculated
these values by
averaging over all components in each epoch. The error bars represent
the standard deviation of average fractional linear polarization.
Over the whole shell, the mean linear polarization rises from
$\phi$ =0.158 until  $\phi$= 0.744 when it reaches the maximum
value. The mean fractional polarization then
drops abruptly from the maximum to reach its
minimum at $\phi$= 0.873.  After that, it tends to increase again till
$\phi$=1.783.

Total intensity, polarized intensity and fractional polarization for
the inner ring (25 mas radius), and for the whole shell, were plotted
with Hanning smoothing over 5 channels, giving $\sim$1 km s$^{-1}$ effective
resolution.

\begin{figure*}
  \includegraphics[width=170mm,height=105mm]{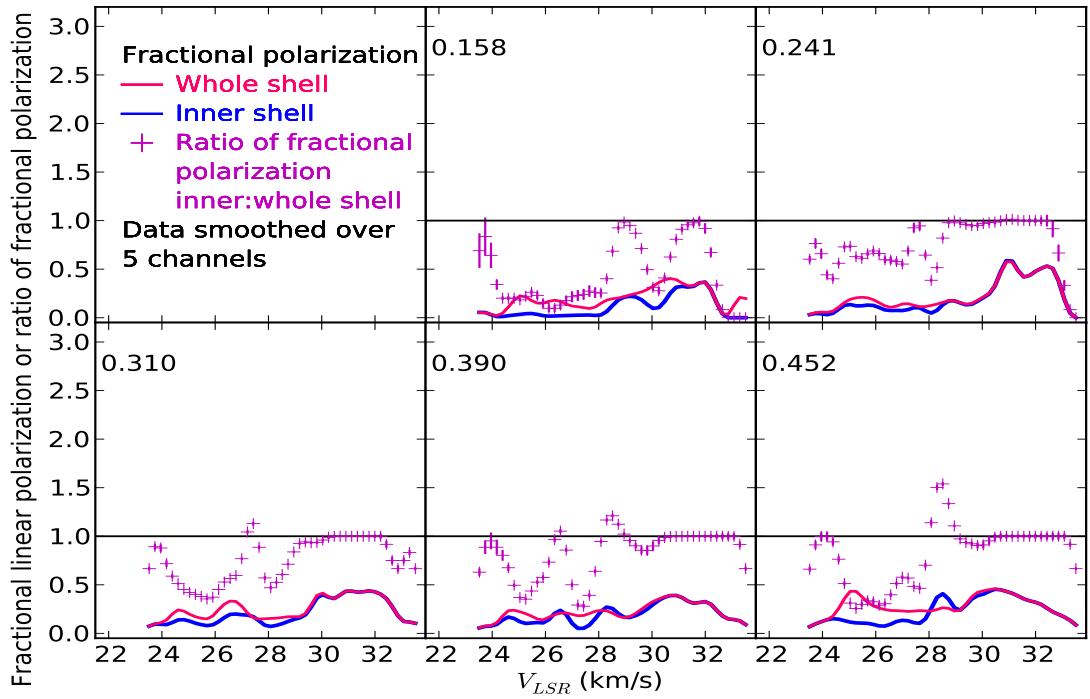}
\caption{Ratio of fractional linear polarization in the inner ring, to the
whole shell for the phases as labelled at epochs 1--5.}
\label{f8}
\end{figure*}

$\bullet$ Fig~ \ref{f8}, (Epochs 0.158--0.452),
The ratio of fractional
polarization in the inner shell to the whole shell barely exceeds
1. The overall flux density is high at these epochs and the fractional
polarization is mostly $<$ 50$\%$, higher at red-shifted velocities where
the emission is almost entirely from the inner shell.  However, at
blue-shifted velocities, where there is more extended emission, this
is usually more polarized. At epoch 0.310 and later, the fractional
polarization in the inner shell is slightly higher than in the whole
shell, at a few central velocities.

\begin{figure*}
  \includegraphics[width=170mm,height=105mm ]{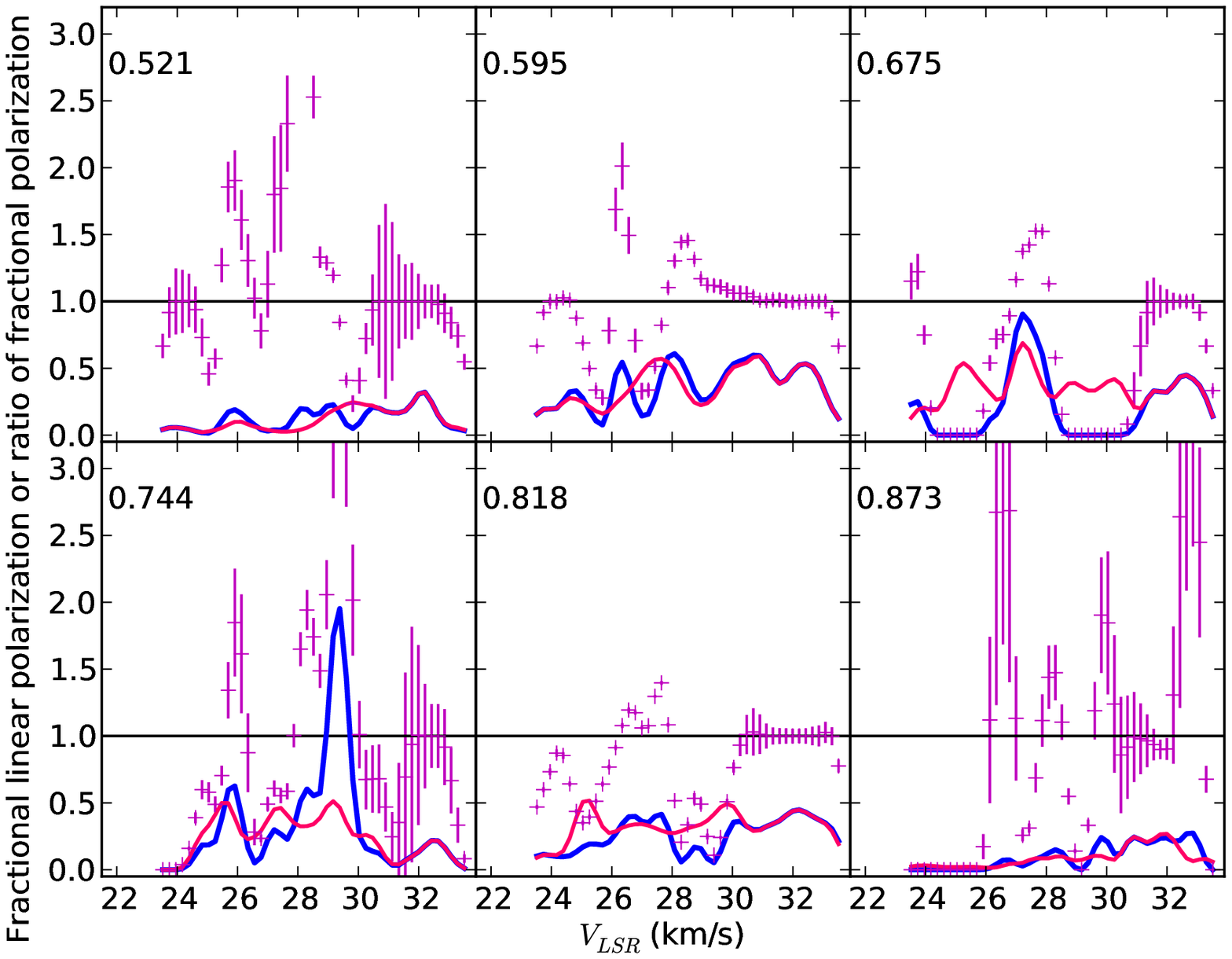}
\caption{Same as Fig \ref{f8} but for epochs 6--11}
\label{f9}
\end{figure*}

$\bullet$ Fig~ \ref{f9}, (Epochs 0.521 -- 0.818),
The overall flux density decreases at all epochs except 0.818. The
fractional polarization is quite variable, and at epoch 0.744 the
fractional polarization from the inner shell exceeds 100$\%$.  The
ratio of fractional polarization in the inner shell to that in the
whole shell also exceeds 1, and although the uncertainties are high
the excess is significant, mostly at intermediate velocities.

\begin{figure*}
  \includegraphics[width=170mm,height=105mm ]{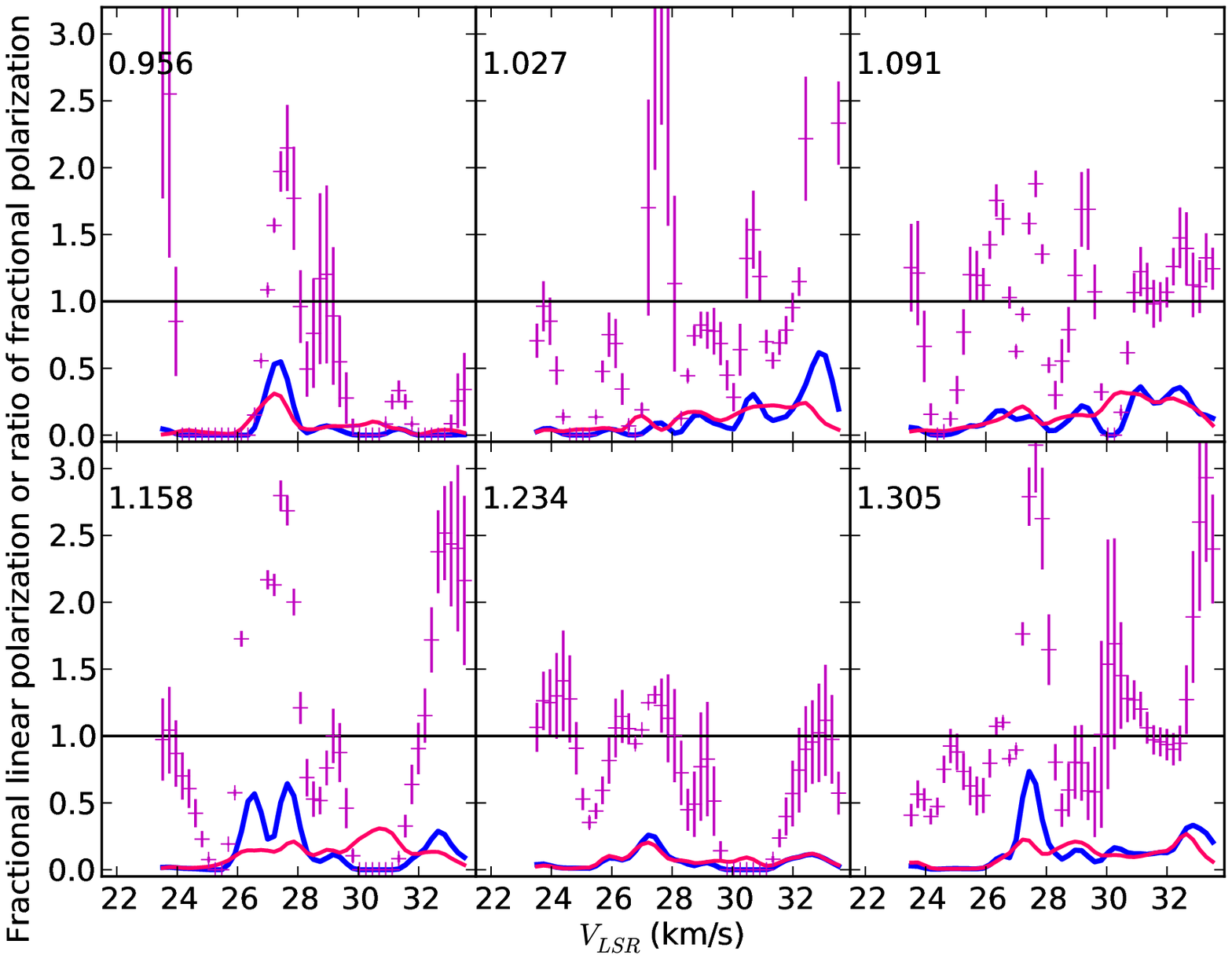}
\caption{Same as Fig \ref{f8} but for epochs 12--17}
\label{f10}
\end{figure*}

$\bullet$ Fig~ \ref{f10}, (Epochs 0.873 -- 1.305),
The trend continues, with higher fractional polarization from the
inner shell at a range of
velocities.  The total intensity from the whole shell increases, but
with fractional polarization mostly $<$ 30$\%$, whereas from the inner shell
it exceeds 50$\%$ at most epochs at some velocities.

\begin{figure*}
 \includegraphics[width=170mm,height=105mm ]{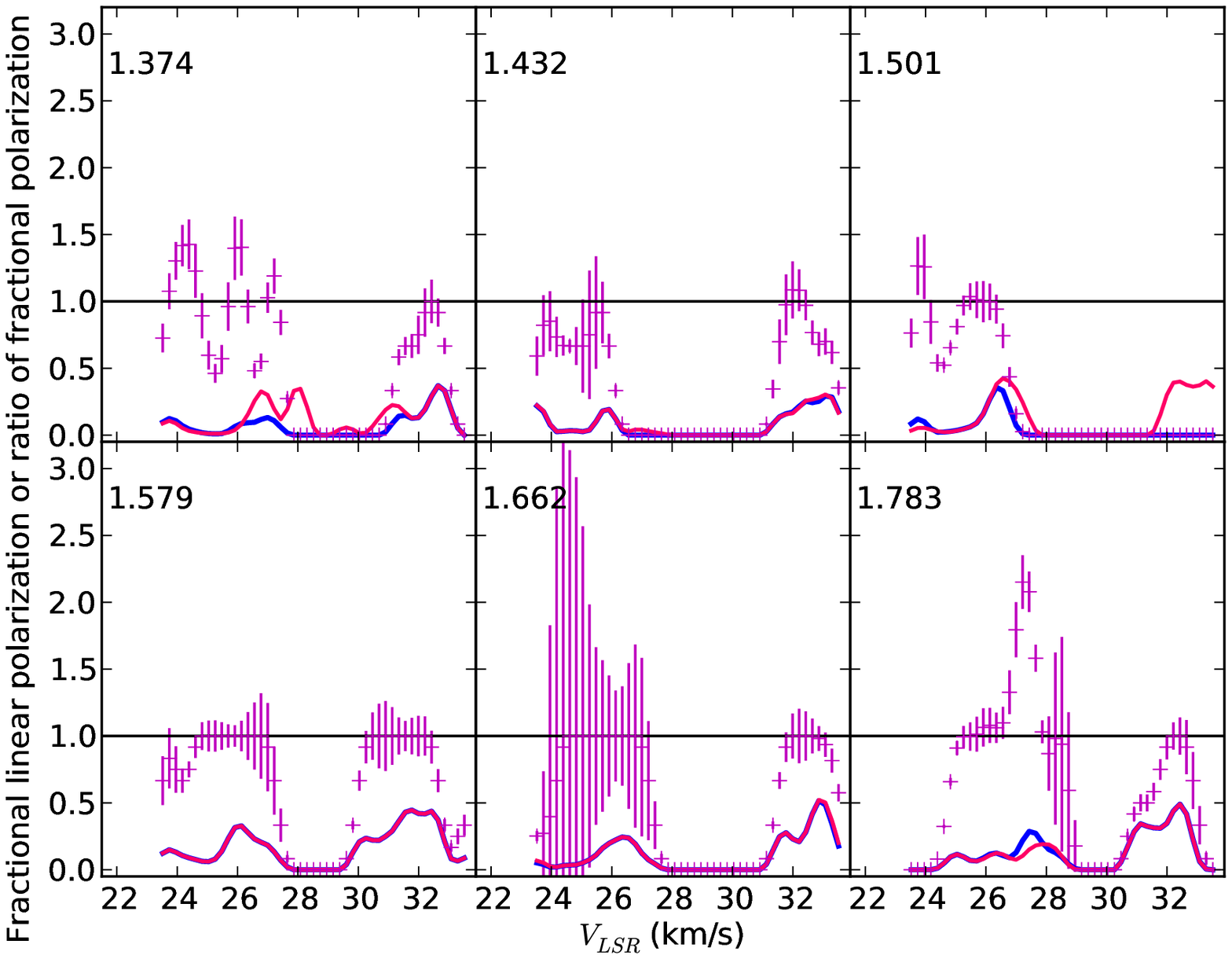}
\caption{same as Fig \ref{f8} but for epochs 18--23}
\label{f11}
\end{figure*}

$\bullet$ Fig~ \ref{f11}, (Epochs 1.374 -- 1.783),
The total intensity decreases, especially at red-shifted velocities,
and the emission is mostly from the inner shell, so the ratio of
polarized intensities is close to 1, although it exceeds 1 at epochs
1.374 and 1.783, when there is more emission from the whole shell but
it is on average less polarized than from the inner shell.

Fig. ~\ref{f6} shows that the fractional polarization tends
to be greater when the total intensity is weaker.  Figs.~\ref{f8} --
\ref{f11}
show that this is often due to one or two narrow spectral regions,
usually near the centre of the spectra, which have high fractional
polarization. These are associated with the inner shell, see $\phi$
0.675, 0.744 and 0.956 in the first cycle.  Similar behaviour is seen
in the second cycle at several epochs.

Thus, there is a distinct tendency for higher fractional polarized
intensity to be associated with weaker total intensity and with the
inner shell.
 The emission at more extreme velocities tends
to come from the inner shell, as projected on the sky.  This is not
surprising since, for a spherical shell with a radial velocity field,
we would expect to see the most red- and blue-shifted emission close
to the line of sight to the star.  Note that  red-shifted emission
within the innermost 25 mas ($R_{\star}$, \citealt{weigelt2000}) must
be on the near side of the star, seen in infall.

There are no obviously cyclic repeating trends in the second cycle
compared with the first.  The first half  of the
first cycle shows lower fractional polarization in the inner shell for
most spectral features whilst the inner shell dominates the
polarization in the second half of the cycle.  In the second cycle,
most of the polarization comes from the inner shell but there is no
clear dependence on phase.

What does this imply?  If a magnetic field is responsible for
polarization, this would be expected to be stronger nearer the star,
and might also be enhanced by shock compression.  Since the shock
crossing time would not necessarily be an exact multiple of the
stellar period, this could produce the out-of-phase enhancement.  If,
on the other hand, the mechanism were radiative, that could also be
strongest in the inner shell but would be expected to follow the
stellar phase, so this is less likely.

\subsection{Alfv\'en speed}

If we use the estimate of $B$ based on equating the  thermal energy
density with the energy density
in the magnetic field, we expect the
Alfv\'en speed $\nu_A$ and sound speed $\nu_s$ to be similar.

We calculated the sound speed by using the relation:
\begin{equation}
\nu_{s}=\sqrt{\frac{\gamma kT}{m}}
\label{E10}
\end{equation}
where $\gamma$ is the adiabatic constant, $k$ is the  Boltzmann constant
and $T$ is the effective temperature in the masing region.
We found the sound speed $\nu_{s}=$ 2.77 km s$^{-1}$.  The total random energy in
the gas is likely to be slightly higher due to other turbulence.

 The Alfv\'en speed $\nu_A$ is given by:
\begin{equation}
\nu_{A}=\frac{B}{\sqrt{\mu_{\circ}\rho}}
\label{E8}
\end{equation}
where $\rho$ is the total mass density for well-coupled fluid and it
is given by
\begin{equation}
\rho=\sum{n_{s}m_{s}}
\label{E9}
\end{equation}
where $\mu_{\circ}$ is the permeability of vacuum ($4\pi\times10^{-7}$) H m$^{-1}$.
From \citet{reid1997} we found the total density is $0.4\times10^{11}$cm$^{-3}$
at the midpoint in the maser shell ($7.3\times10^{13}$ cm). We found
$\nu_A=$3.5 km s$^{-1}$ using $B=750$ mG, or higher if the values of
$B$ from \citet{herpin2006} are considered.  Thus, in the SiO maser
clumps, $\nu_A$ and $\nu_s$ are similar.  In a more diffuse,
highly-ionised region, with lower density and poorer coupling, $\nu_A$
could be considerably higher.

\section{Conclusions}{\label{s5}}

The SiO maser emission is significantly linearly polarized.
Most of
the linear polarization vectors are either perpendicular or tangential
to the projected shell, but other angles can be seen.
During the first cycle, up to phase 1.158, two-thirds of epochs have
the majority of the emission with polarization angles parallel to
radial direction, i.e. parallel to the direction of outflow of the
first cycle. Emission with perpendicular position angles is commonest in the
other five epochs. Emission with intermediate polarization position
angles dominates the remaining epochs, but the intensity tends to be
lower leading to greater uncertainties.

The bimodal distribution of {\sc evpa} is consistent with that reported by
\citet{cotton2006} in VLBA observations of SiO masers in Mira variable
stars at 7 mm. They reported that for some features the {\sc evpa} are
tangential while for the others are radial. The  linear polarization
{\sc evpa} is predominantly tangential to the projected intensity shell in
the VLBA observation toward TX Cam \citep{kemball2009}. This pattern has
been confirmed in further observations toward TX Cam and  IRC+10011
\citep{desmurs2000}. However, \citet{cotton2004} found there was no prevalent
pattern of {\sc evpa} in any observed stars.

\begin{figure}
\begin{center}
    \includegraphics[width=85mm]{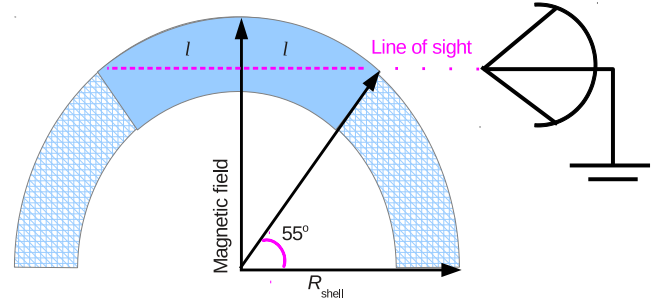}
\end{center}
\caption{Cartoon showing an observation of a maser shell with a radial magnetic
field.  The angle between the magnetic field and the line of sight
$\theta_{\mathrm F}$ is
$<55^{\circ}$ in the hatched area, producing {\sc evpa} parallel to
the magnetic field; in the solidly shaded area the angle is
$>55^{\circ}$ and the  {\sc evpa} is perpendicular. At the transition
point the maximum line of sight through the shell is $2 l$.}
 \label{f15}
\end{figure}

Our data are in the limit g$\Omega \gg$ R $\gg\Gamma$, which indicates
that the \citet{goldreich1973} (GKK) solutions for the linear
polarization are applicable, i.e.  the standard interpretation for the
radially and tangentially {\sc evpa} orientations relative to
projected magnetic field is applicable.  This result is consistent
with that reported by \citet{kemball2009} toward TX Cam. The high
brightness temperatures observed indicate that the R Cas masers are
predominantly saturated so we consider the GKK model for saturated
emission.  We used the GKK model to estimate the depth, $l$, above
the midplane of the shell, of the feature showing an abrupt
90$^{\circ}$ polarization angle change, see Fig~\ref{f15}.  At the
point where the (dashed) line of sight crosses the vertical magnetic
field line, in the plane of the sky, the angle between the magnetic
field direction and the line of sight, $\theta_{\mathrm F}$, is
$90^{\circ}$ and $0^{\circ}\le$ {\sc evpa} $\le35^{\circ}$ with
respect to an axis perpendicular to the local magnetic field
direction.  When $\theta_{\mathrm F}$ reaches 55$^{\circ}$ (the van
Vleck angle), the {\sc evpa} becomes parallel to the direction of the
magnetic field as projected against the plane of the sky.  At this
point, Eq.~\ref{E11} gives $l$ equivalent to $\sim 15$ mas or 2.6 AU.

\begin{equation}
\cos55=\frac{l}{R_{\mathrm{shell}}}
\label{E11}.
\end{equation}
 
If the magnetic field is radial, then the preponderance of {\sc evpa} parallel to the radial direction (during the first cycle, with the best quality data,  Section~\ref{sec:linpol}) suggests that a substantial fraction of the masers which we detect come from  a region at least $\pm 2.6$ AU deep with respect to the plane
of the sky containing the star. This suggests a shell thickness of $\ge4.6$ mas or 0.8 AU, which is indeed similar to or less than the values of d$r$ given in Table 1. The shell thickness varies from about 1/5 to 2/3 of the shell radius, error-weighted mean 1/3. The next most common orientation of {\sc evpa} is tangential, suggesting that it is close to perpendicular to the magnetic field, when emanating from masers closer to the plane of the sky.
The magnetic field lines must close at larger
distances or orientations.

The percentage of the linear polarization is m$_{\ell}\sim$ 11$\to$
58$\%$ averaged over features for each epoch. The highest fractional
polarization, from the inner shell, is seen in some features in epochs
where the total intensity is lower. There is a
tendency at all epochs for the fractional polarization in the inner
shell to be greater at the line peaks. Figs.~\ref{f8} to~\ref{f11}  do not show any other  relationships between
fractional polarization and velocity.  Neither is
there any discernible correlation between $\theta_{\mathrm F}$ and velocity.

The changes in the fractional linear polarization strength are
probably due to fluctuations in the maser amplification process rather
than the magnetic field strength. The changes in polarization vector
orientation are likely to be due to small variations in the magnetic
field direction which is possibly caused by local turbulence (see
Section~\ref{Zeint}). Our estimates of the magnetic field strength in the maser region
make pumping models based on radiation-driven anisotropies in the
magnetic sublevel populations highly unlikely.

A few isolated features with fractional polarization apparently
exceeding 100\% can be explained if there is structure in the ionised
fraction or the magnetic field on scales $<$5 mas but the total
intensity emission is smooth on larger scales and a greater proportion
is resolved out by the VLBA. Since the Alfv\'{e}n speed is unlikely to
exceed the sound speed significantly, and disturbances at either
velocity would take $\ga1$ yr to cross a 5-mas clump, intrinsic
variability is unlikely.  However, Faraday rotation is not negligible
at 7 mm wavelength.  We found that it is $\sim15^{\circ}$ for masers
propagating through the depth of the SiO shell, using our estimate of
the magnetic field strength $B=0.725$ G based on the energy balance,
or slightly higher if considering the values of $B$ measured by
\citet{herpin2006}.  If there are small-scales inhomogeneities in the
ionised fraction of the stellar wind, this will affect the propagation
of the polarized emission.  The small-scale structure seems to be
imposed on the maser polarization between the emitting material and
the observer but the effect must occur close to the star where the
magnetic field is strong enough.  This mechanism is also likely
to affect other maser features with high observed fractional linear
polarization, such that although the measured $m_{\ell}$=11--58 percent, the
higher values might be approximately halved if all the emission was
measured by the interferometer.  We are, therefore, probably within
the limit of the GKK model, wherein the mean fractional linear
polarization can be up to 33 percent.

\section*{Acknowledgments}
The authors thank the VLBA for providing the data used in this
paper. KAA would like to thank the Iraqi government for giving him the
opportunity to gain his PhD at the University of Manchester. We
thank the referee for comments which led to significant improvements
in explanations and in the information conveyed by the figures.




\label{lastpage}

\end{document}